\documentstyle[epsf,prb,aps,floats,twocolumn]{revtex}

\newcommand{\simj}{\stackrel{>}{_\sim}}
\newcommand{\simk}{\stackrel{<}{_\sim}}

\begin{document} 
 \draft
\wideabs{ 

\title{Critical behavior near the metal-insulator transition in the
 one-dimensional extended Hubbard model at quarter filling}

\author{K.  Sano}
\address{Department of Physics Engineering,  Mie University, Tsu, Mie
 514-8507 
, Japan} 
\author{Y. \=Ono}
\address{Department of Physics, Niigata University, Ikarashi, Niigata
 950-2181
, Japan}

\date{\today}
\maketitle

\begin{abstract}
We examine the critical behavior  near the metal-insulator transition (MIT)
 in the one-dimensional extended Hubbard model with the on-site and the
 nearest-neighbor interactions $U$ and $V$ at quarter filling using a
 combined method of the numerical  diagonalization and the  renormalization
 group (RG). 
The Luttinger-liquid parameter $K_{\rho}$ is calculated with the exact
 diagonalization for finite size systems and is substituted into the RG
 equation as an initial condition to obtain $K_{\rho}$ in the infinite size
 system. This approach also yields the charge gap $\Delta$ in the insulating
 state near the MIT. 
The results agree very well with the available exact results for $U=\infty$
 even in the critical regime of the MIT where the characteristic energy
 becomes exponentially small and the usual finite size scaling is not
 applicable. 
When the system approaches the MIT critical point $V\to V_c$ for a fixed $U$, 
$K_{\rho}$ and $\Delta$ behave as  
$|\ln \Delta|^{-2}=c_\Delta(V/V_c-1)$ and 
$(K_{\rho}-\frac{1}{4})^2=c_K(1-V/V_c)$,
where the critical value $V_c$ and the coefficients $c_\Delta$ and $c_K$ are
functions of $U$. 
These critical properties, which are known to be exact for $U=\infty$, are
 observed also for finite $U$ case. 
We also observe the same critical behavior in the limit of the MIT critical 
point $U\to U_c$ when $U$ is varied for a fixed $V$. 
\end{abstract} 
%\kword{Metal-Insulator transition; Extended Hubbard Model; Numerical 
% Diagonalization; Renormalization Group Method}
}

\section{Introduction}
A number of theoretical studies have been made on the one-dimensional (1D)
 extended Hubbard  model with the on-site interaction $U$ and the
 nearest-neighbor interaction $V$ as a simple model for quasi-1D 
 materials \cite{Emery1,Solyom,Voit,Hirsch}. 
It has been reported that this model shows a rich phase diagram including 
the metal-insulator transition (MIT), the phase separation, the spin-gapped 
phase and the superconducting (SC) phase \cite{Mila,Sano1,Penc}. 
In particular, the MIT at quarter-filling has attracted much interest as it 
takes place for finite values of $U$ and $V$ in contrast to the MIT at 
half-filling where the system is insulating except for $U=0$. 
Therefore, the MIT at quarter-filling is important as a typical example of the 
quantum phase transition caused by the electron correlation, 
and have been extensively studied by 
many authors \cite{Penc,Nakamura,Sano2,Sano3,Ohta,Tsuchiizu,Yoshioka}. 
However, the critical properties of the MIT have not received so much 
attention as it is more difficult to investigate the system in the limit of 
the MIT where the characteristic energy  becomes exponentially small. 
In this paper, we wish to study the critical behavior near the MIT 
which is a typical example of the quantum critical phenomena caused by the 
electron correlation.

The extended Hubbard  model is given by the following Hamiltonian
\begin{eqnarray} 
H&=&-t\sum_{i,\sigma} (c_{i\sigma}^{\dagger} c_{i+1\sigma}+h.c.)
        \nonumber \\
    &+&  U\sum_{i}n_{i\uparrow}n_{i\downarrow}
    +V\sum_{i,\sigma \sigma' }n_{i\sigma}n_{i+1\sigma'}, 
\end{eqnarray} 
where $c^{\dagger}_{i\sigma}$  stands for the creation operator of an
 electron with spin $\sigma$  at site $i$ and
 $n_{i\sigma}=c_{i\sigma}^{\dagger}c_{i\sigma}$.
$t$ represents the transfer energy between the nearest-neighbor sites and is
 set to be unity ($t$=1) in the present study.
It is well known that this Hamiltonian (1) can be mapped on an $XXZ$ quantum
 spin Hamiltonian in the limit $U\rightarrow\infty$. 
The term of the nearest-neighbor interaction $V$ corresponds to 
 the $Z$-component of the antiferromagnetic coupling   and the transfer
 energy $t$ does  the $X$-component. 
When the $Z$-component is larger than the $X$-component, the system has a
 "{\it Ising}"-like symmetry and an excitation gap exists.
For the  Hubbard  model, this corresponds to the case with $V>2t$ 
 where the charge gap is  exactly obtained \cite{Yang}. 
On the other hand, in the case with "$XY$"-like symmetry ($V<2t$),
the system is metallic and  the Luttinger-liquid parameter $K_\rho$ is
 exactly given by $\cos(\frac{\pi}{4K_{\rho}})=-V/2$\cite{Luther}.

In the finite $U$ case,   exact results have not been obtained  except for
 $V=0$. 
In this case, the weak coupling renormalization group method (known as 
 $g$-ology) and the exact (numerical) diagonalization (ED) method have been  
 applied.
The  $g$-ology yields the phase diagram of the   1D extended Hubbard  model 
  analytically, but  quantitative validity is guaranteed only in the weak
 coupling regime \cite{Emery1,Solyom,Tsuchiizu,Yoshioka}. 
On the other hand, the numerical approach is a useful method to examine 
 properties of the model in the strong coupling regime
 \cite{Hirsch,Mila,Sano1,Penc,Nakamura,Sano2,Sano3,Ohta}.
In particular, the numerical diagonalization of a finite-size system has
 supplied us  with reliable and important information
 \cite{Mila,Sano1,Penc,Nakamura,Sano2,Sano3}. 

However, it is difficult for purely numerical approaches to investigate 
 the critical behavior near the MIT where the characteristic energy scale 
 of the system becomes exponentially small. 
To overcome this difficulty, we have recently proposed a combined approach 
of the ED and the RG methods. 
\cite{Sano2,Sano3}.
This approach enables us  to obtain accurate results of  the Luttinger-liquid 
 parameter  $K_\rho$ and the charge gap $\Delta$  near the MIT  beyond 
 the usual finite size scaling for the ED method. 
The obtained results of $K_\rho$ and $\Delta$ have been compared with 
the available exact results for $U=\infty$ and found to be in good 
agreement \cite{Sano3}. 
The phase diagram of the MIT at quarter-filling together with the contour map 
of the charge gap $\Delta$ has been obtained on the $U$-$V$ plane \cite{Sano3}. 
However, the critical behavior near the MIT was not discussed 
in the previous work. Here we extensively apply 
this approach to the critical regime of the MIT to elucidate the critical 
behavior of $K_\rho$ and $\Delta$ in the limit of the MIT.

\section{Luttinger liquid  and RG method}
 First, we briefly discuss a general argument for 1D-electron systems
 based on the  bosonization theory \cite{Emery1,Solyom,Voit}.
According to this theory, the effective Hamiltonian  can be separated into
 the charge  and spin parts. 
So, we turn our attention to only the charge part and do not consider the
 spin part in this work.
In the low energy limit, the effective Hamiltonian of the charge part  is 
 given by
\begin{eqnarray} 
     H_{\rho}&=&\frac{v_{\rho}}{2\pi}\int_0^L {\rm d}x
  \left[K_{\rho}(\partial_x \theta_{\rho})^2
       +K_{\rho}^{-1}(\partial_x \phi_{\rho})^2\right]   \nonumber
\\
   &+& \frac{2 g_{3\perp}}{(2\pi\alpha)^2}
  \int_0^L {\rm d}x \cos[\sqrt{8}\phi_{\rho}(x)]
\end{eqnarray}
where $v_{\rho}$ and $K_{\rho}$ are the charge velocity and the coupling 
 parameter, respectively. 
The  operator $\phi_{\rho}$ and the dual operator $\theta_{\rho}$ represent
 the phase fields of the charge part. 
 $g_{3\perp}$   denotes the amplitude of the umklapp  scattering and
 $\alpha$ is a short-distance cutoff.  

In the Luttinger liquid theory,  some relations  have been  established as
 universal relations in one-dimensional  models.\cite{Voit}  
 In  the model which is isotropic in spin space,  the critical exponents of
 various types of correlation functions are determined by a single parameter
 $K_{\rho}$.
  It is  predicted that the SC correlation  is dominant
 for $K_{\rho}>1$ (the correlation function decays as $\sim
 r^{-(1+\frac{1}{K_{\rho}})}$), whereas  the CDW  or  SDW correlations are
 dominant  for $K_{\rho}<1$ (the correlation functions decay  as $\sim
 r^{-(1+K_{\rho})}$)  in the Tomonaga-Luttinger liquid \cite{Voit}.
The critical exponent $K_{\rho}$ is related to  the charge susceptibility
 $\chi_c$ and  the Drude weight $D$ by
\begin{equation}
      K\sb{\rho}=\frac{1}{2}(\pi \chi_c D)^{1/2},
\end{equation}
with $D=\frac{\pi}{N_a} \frac{\partial^2 E_0(\phi)}{\partial \phi^2}$, where
 $E_0(\phi)$ is the total energy of the ground state as a function of  a
 magnetic flux $N_a \phi$ and  $N_a$ is  the system size \cite{Voit}. 
Here, the magnetic flux is imposed by  introducing the following  gauge
 transformation:  $c_{m\sigma}^{\dagger} \to
 e^{im\phi/N_a}c_{m\sigma}^{\dagger}$ for an arbitrary site $m$.

When the charge gap vanishes in the thermodynamic limit, the uniform  charge
 susceptibility $\chi_c$ is obtained from 
\begin{equation}
\chi_c=\frac{4/N_a}{E_{0}(N_{e}+2,N_a)+E_{0}(N_{e}-2,N_a)-2E_{0}(N_{e},N_a)},
\end{equation}
where $E_{0}(N_e,N_a)$ is the  ground state energy of a system with $N_a$
 sites and $N_e$ electrons. Here, the filling $n$ is defined  by 
 $n=N_{e}/N_{a}$.
 
We numerically diagonalize the Hamiltonian Eq. (1)  up to 20 sites system
 using the standard Lanczos algorithm. Using the definitions of Eqs. (3) and
 (4), we calculate  $D$ and $\chi_c$ from the ground state energy of the
 finite size system.
To carry out a systematic calculation, we use the periodic boundary
 condition for $N_e=4m+2$ and the antiperiodic boundary condition for
 $N_e=4m$, where $N_e$ is the total electron number and $m$ is  an  integer.
 This choice of the boundary condition removes accidental degeneracy so that
 the ground state might always be a singlet with zero momentum.

%*******************************************************************************
\begin{figure}[t]
  \begin{center}
\epsfxsize=7cm
\epsffile{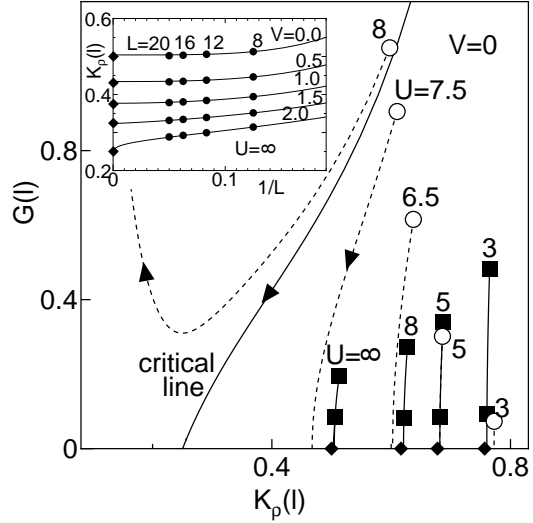}
\end{center}

  \caption[]{
    The RG flow obtained from  the numerical initial condition (solid lines)
 and that from the analytical one (broken lines) for various $U$ at $V=0$.
 The filled squares are the numerical initial conditions with $L_1=8$ and
 $L_2=12$, and the open circles are the analytical ones. 
The filled diamonds on the $K_{\rho}$ axis are the exact results for $U=$3,
 5, 8 and $\infty$. 
Inset shows the RG flow of $K_{\rho}(\ell)$ (solid lines), the numerical
 result of $K_{\rho}(\ell)$ (filled circles) and the exact result (filled
 diamonds) for various $V$ at $U=\infty$. 
}
  \label{fig:1}
\end{figure}
%*******************************************************************************

At quarter filling, the $8k_F$ umklapp scattering is crucial to
 understanding  the MIT. 
 The effect of the umklapp term is renormalized   under the change of the
 cutoff $\alpha\rightarrow {\rm e}^{\ell}\alpha$.
In this work, we adopt the Kehrein's formulation as the RG equations
 \cite{Kehrein1,Kehrein2} 

\begin{eqnarray}
 \frac{{\rm d}K_{\rho}(\ell)}{{\rm
 d}\ell}&=&-8\frac{G(\ell)^2K_{\rho}(\ell)^2}{\Gamma(8K_{\rho}(\ell)-1)},\\
 \frac{{\rm d}\log G(\ell)}{{\rm d}\ell}&=&[2-8K_{\rho}(\ell)],
\end{eqnarray}
where  the scaling quantity $\ell$ is related to the cutoff $\alpha$, 
 $\Gamma(x)$ is $\Gamma$-function and  $G(0)=g_{3\perp}/2\pi \alpha^2
 v_{\rho}$.
 This  formulation is an extension of the perturbative RG theory and 
allows us to estimate  the charge gap together with $K_{\rho}$ in the
 infinite size system. 
To solve these equations concretely, we need an initial condition for the
 two values: $K_{\rho}(0)$ and $G(0)$.
Here, the value of the short-distance cutoff $\alpha$ is selected to  a
 lattice constant of the system and set to be unity.
Although the continuum field theory does not give  this  cutoff
 parameter, our choice  is quite natural to apply this method to the lattice
 system.

In the weak-coupling limit,  analytic expressions for the initial condition
 have been obtained \cite{Yoshioka}. 
 At quarter filling, $v_{\rho}$, $g_{3\perp}$ and $K_{\rho}(0)$ are given by
 $\{(2\pi v_F +U+4V)^2-(U+4V)^2\}^{1/2}/2\pi$, $(U-4V)U^2/(2\pi v_F)^2$ and 
 $\{1+(U+4V)/(\pi v_F)\}^{-1/2}$, respectively, where $v_F$ is  $2t\sin
 k_F$.
When we substitute these values into the RG equations as the initial
 condition, we find  the insulating states  for  $U\simj 8$ at $V=0$ in
 contrast to the exact results which show the metallic states for all $U$ at
 $V=0$. 
 This inconsistency suggests  that  the  analytical initial condition  is
 not applicable  in the strong coupling regime. 

To find an adaptable initial condition to the strong coupling regime, we
 diagonalize a $L$-site system numerically and calculate $K_{\rho}(\ell)$ by
 using the relation $\ell \simeq \ln L$. 
It is  easy for the numerical calculation to obtain $K_{\rho}(\ell)$ as
 compared to $G(\ell)$. 
Therefore, we    calculate  $K_{\rho}(\ell_1)$ and $K_{\rho}(\ell_2)$ with
 $L_1$- and $L_2$-site systems instead of $K_{\rho}(\ell)$ and $G(\ell)$
 with a $L$-site system.
To eliminate $G(\ell)$ in the RG equations, we  integrate  the Eq. (6) 
and obtain, 
\begin{equation}
G(\ell)=G(\ell_0)e^{\int_{\ell_0}^{\ell}[2-8K_{\rho}(\ell')]d\ell'}, 
\label{Gell}
\end{equation}
where  $\ell_0$ is a constant.
Then the differential equation for $K_{\rho}(\ell)$  is written by
\begin{equation}
\frac{{\rm d}K_{\rho}(\ell)}{{\rm d}\ell}=-8\frac{G^2(\ell_0)
 e^{\int_{\ell_0}^{\ell}[4-16K_{\rho}(\ell')]d\ell'}
 K_{\rho}(\ell)^2}{\Gamma(8K_{\rho}(\ell)-1)}.
\end{equation}
When we set $\ell_0=\ell_1$ and use $K_{\rho}(\ell_1)$  as the initial
 condition for the above   equation, we can   obtain  the solution 
 numerically except the constant $G(\ell_1)$.
By fitting   the value of this solution  at $\ell =\ell_2$ to 
 $K_{\rho}(\ell_2)$, we  can determine   $G(\ell_1)$. Then,  $G(\ell)$ is
 immediately calculated from Eq. (7) and  the solution of the RG equations is
 completely  obtained.
% 

%*******************************************************************************
\begin{figure}[t]
  \begin{center}
\epsfxsize=7cm
\epsffile{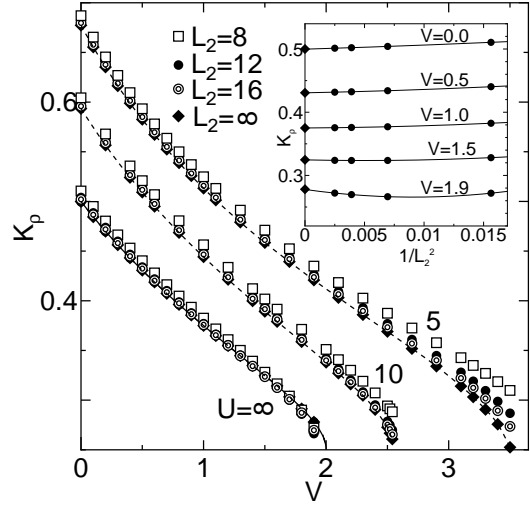}
\end{center}
  \caption[]{
 $K_{\rho}^{L_2}(\infty)$ as a function of $V$ at $U=5$ $10$ and $\infty$ 
with the exact result at $U=\infty$ (the solid line). The broken lines are
 guides for the eye for the data of 
$K_{\rho}^{\infty}(\infty)$ .
Inset shows $K_{\rho}^{L_2}(\infty)$ as a function of $1/L_2^2$  for  
 various $V$ at $U = \infty $ (filled circles) together with the exact
 result (filled  diamonds). 
 }
\label{fig:2}
\end{figure}
%*******************************************************************************

\section{detailed analysis of $K_\rho$ }

In Fig. 1, we   show the RG flow obtained by solving the  RG equations  with
 the numerical  and the analytical initial conditions  on the
 $K_{\rho}(\ell)$-$G(\ell)$ plane. 
Here, we set $L_1=8$ and $L_2=12$ for the numerical initial condition. 
 In the weak coupling regime with $U\simk 5$,  the renormalized 
 $K_{\rho}(\infty)$ obtained from both the initial conditions agree with the
 Bethe ansatz result. 
 On the other hand, in the strong coupling regime with $U\simj 5$, there is
 a large discrepancy between $K_{\rho}(\infty)$ with the analytical initial
 condition and the exact result. This is a striking contrast to the
 numerical initial condition which yields $K_{\rho}(\infty)$ in excellent
 agreement with the exact result even in the limit $U\rightarrow\infty$. 

In the inset of Fig. 1, we plot the RG flow of $K_{\rho}(\ell)$  as a
 function of  $L^{-1}$ for various $V$ at $U=\infty$   together with
 $K_{\rho}(\ell)$ from the numerical diagonalization for several system
 sizes $L$ and from the exact results for $L=\infty$. 
The RG flow seems to  connect smoothly the numerical results and the exact
 result.  It indicates that the size dependence of $K_{\rho}(\ell)$ is well
 described by the RG equations.  
For $U=\infty$, our result is consistent with the previous result from Emery
 and  Noguera \cite{Emery-Noguera}. They solved the RG equations with the
 numerical initial conditions $K_{\rho}(\ell)$ and $G(\ell)$ for a system
 size $L$,  
where  $G(\ell)$ is calculated from the excited state energy. On the other
 hand, in our approach, we need only the value of $K_{\rho}(\ell)$ which is
 calculated from the ground state. Then, our approach can be easily extended
 to a complicated model such as the extended Hubbard model with finite $U$
 in contrast to the previous approach \cite{Emery-Noguera} which has been
 applied only for the infinite $U$ case.

%*******************************************************************************
\begin{figure}[t]
  \begin{center}
\epsfxsize=7.8cm
\epsffile{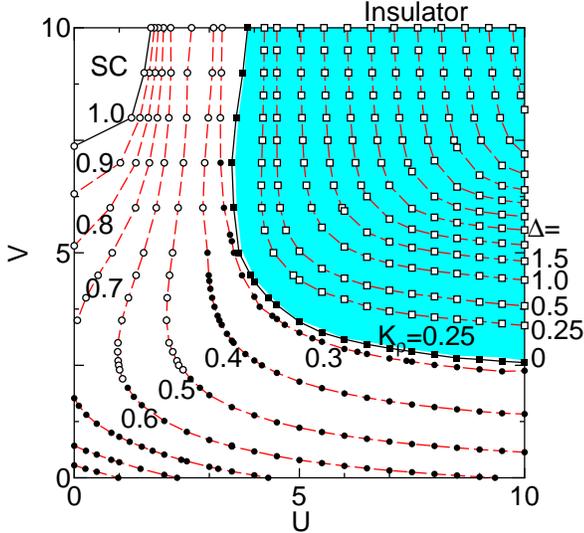}
\end{center}
    \caption[]{
The contour lines for $K_{\rho}^\infty(\infty)$  and for the charge gap  on
 the $U$-$V$ plane.
The sold circles are $K_{\rho}^\infty(\infty)$ from the RG equations with
 numerical initial conditions, and the open  circles are  the numerical
 diagonalization results (see in the text). 
The open squares represent the charge gap (Ref. 10), whose values are 0.25, 0.5, 1, 1.5, 2, 2.5, 3, 3.5, 4, 4.5, 5 and 5.5. 
The filled  squares represent the phase boundary of the MIT where 
 $K_{\rho}^\infty(\infty)=1/4$.
}
\label{fig:3}
\end{figure}

In order to check  the dependence of $K_{\rho}(\infty)$ on the system sizes
 $L_1$ and $L_2$ for the initial condition, we calculate $K_{\rho}(\infty)$ 
 by using the  three different sizes $L_2=8$, $12$ and $16$ with
 $L_1=L_2-4$. 
In Fig. 2, we show  $K_{\rho}(\infty)$ as a function of $V$ for $U=5$, $10$
 and $\infty$ together with the  exact result for $U=\infty$.
The value of $K_{\rho}(\infty)$ is  slightly dependent on $L_2$.
The inset in Fig. 2 shows the $L_2$ dependence of $K_{\rho}(\infty)$  for
 various $V$ at $U= \infty $ together with the corresponding exact result. 
Here, we assume that the size dependence of $K_{\rho}(\infty)$ is given by
 $K_{\rho}^{L_2}(\infty) \sim K_{\rho}^\infty(\infty)+ c_1/L_2^2 +
 c_2/L_2^4$, where  $c_1$ and $c_2$ are constants, and
 $K_{\rho}^\infty(\infty)$  is the $L_2 \to \infty$ extrapolated value  of
 $K_{\rho}(\infty)$.
We see that  $K_{\rho}^\infty(\infty)$ is very close to the exact result for
 $U=\infty$. 
We may expect that the RG equations with numerical initial conditions give a
 reliable estimate for $K_{\rho}^\infty(\infty)$ not only for the infinite
 $U$ case but also for the finite $U$ case where the exact result is not
 known so far. 
%*******************************************************************************
\begin{figure}[t]
  \begin{center}
\epsfxsize=7.5cm
\epsffile{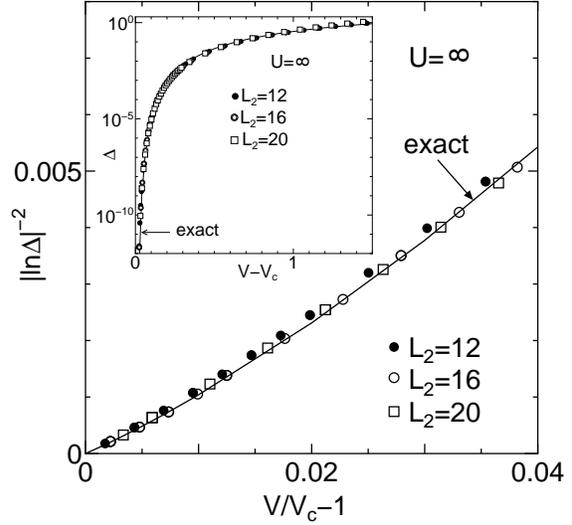}
\end{center}    
\caption[]{$|\ln \Delta|^{-2}$ as a
 function of $(V/V_c-1)$  near the MIT  at $U=\infty$ for $L_2=12$, 16
 and 20  together with the exact result (solid line). 
Inset shows  $\Delta$ as a function of $V-V_c$.}
\label{fig:4}
\end{figure}
%*******************************************************************************

When the strength of $V$ exceeds a critical value $V_c$,  $K_{\rho}(\infty)$
  is renormalized to the value of the strong coupling limit: $K_{\rho}=1/4$.
 This critical point  corresponds to the  MIT point of the system. In the
 infinite $U$ case, we find $V_c \simeq 1.93, 1.95$  and $1.96$ for
 $L_2=12,16$ and $20$, respectively. 
Assuming the size dependence of $V_c$ to be $\propto 1/L_2^2$, 
we  obtain an $L_2 \to \infty$ extrapolated value $V_c=1.99$. 
It  agrees well with the exact value $V_c=2$ for $U=\infty$. 
The similar extrapolation  yields the critical 
 values of the MIT: $V_c \simeq 3.45$ for $U=5$ and $V_c \simeq 2.55$ for
 $U=10$ as shown in Fig. 2 \cite{critical-Vc}. 
The results are in good agreement  with the phase boundary of the MIT in the
 previous works \cite{Mila,Sano1,Penc,Nakamura,Sano2,Sano3}. 
Then it confirms that the combined approach of the ED and  the RG methods 
gives accurate results of $K_{\rho}$  even near the MIT. 

%*******************************************************************************
\begin{figure}[t]
  \begin{center}
\epsfxsize=7.5cm
\epsffile{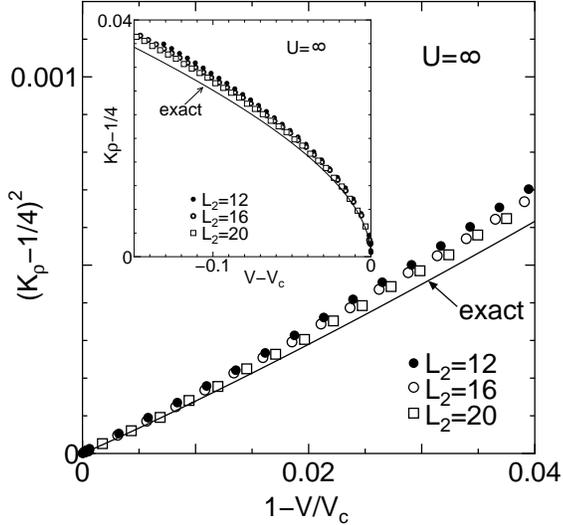}
\end{center}
\caption[]{ $(K_{\rho}-1/4)^2$ as a function of $(1-V/V_c)$  near the MIT 
 at $U=\infty$ for $L_2=12$, $16$ and $20$ together with  the exact result
 (solid line). 
Inset shows  $K_{\rho}-1/4$  as a function of $V-V_c$. 
}
\label{fig:5}
\end{figure}
%*******************************************************************************

In Fig. 3, we show the phase diagram of the MIT on the $U$-$V$ plane together 
 with the contour lines for $K_{\rho}^\infty(\infty)$ in the metallic region. 
 We also plotted the contour lines for the charge gap in the insulating region 
 which have already been reported in our previous paper \cite{Sano3,charge-gap}. 
When $V\gg U$,  the SC phase with $K_{\rho}>1$ appears.
 The character of this phase  has already been discussed in the previous
 works \cite{Mila,Sano1,Penc,Nakamura}. 
Near the SC phase, $K_{\rho}(\ell_2)$ is larger than $K_{\rho}(\ell_1)$ 
 for available finite size systems and, then, we could not obtain the 
 solution of the RG equations for these initial conditions. 
Because the umklapp scattering is canceled by the SC fluctuation, the RG
 equations may not be  applicable in this region.
Thus we estimate  $K_{\rho}$ for $V\gg U$ directly by the ED  method without 
the use of the RG method.

%*******************************************************************************
\begin{figure}[t]
  \begin{center}
\epsfxsize=7cm
\epsffile{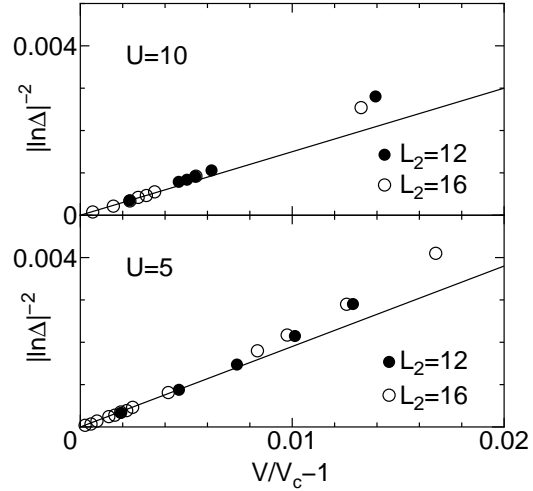}
\end{center}
  \caption[]{
$|\ln\Delta|^{-2}$ as a function of $(V/V_c-1)$ near the MIT  at $U=10$
 and $U=5$ with $L_2=12$  and $16$  together with the straight lines whose
 slope are $0.20$ for $U=10$ and $0.22$ for $U=5$.
Here, the factor of  $\Delta$ is determined by  fitting  $\Delta$ to the
 numerical result at  $V-V_c=1$. 
}
\label{fig:6}
\end{figure}
%*******************************************************************************

%*******************************************************************************
\begin{figure}[t]
  \begin{center}
\epsfxsize=7.15cm
\epsffile{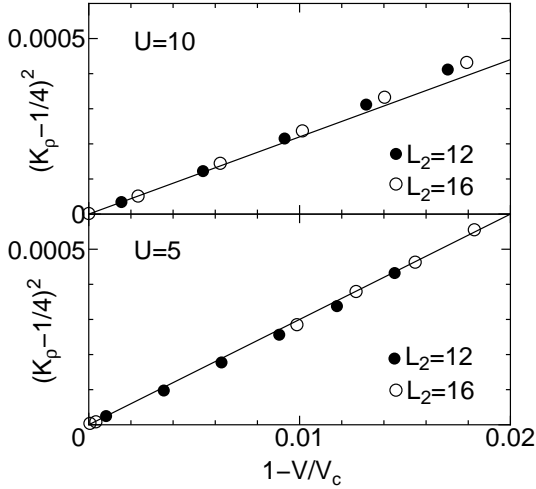}
\end{center}
  \caption[]{
$(K_{\rho}-1/4)^2$ as a function of $(1-V/V_c)$  near the MIT at $U=10$
 and $U=5$ for $L_2=12$  and $16$  together with the straight lines whose
 slope are $0.019$ at $U=10$ and $0.032$ at $U=5$.
}
\label{fig:7}
\end{figure}
%*******************************************************************************

%*******************************************************************************
\begin{figure}[t]
  \begin{center}
\epsfxsize=7cm
\epsffile{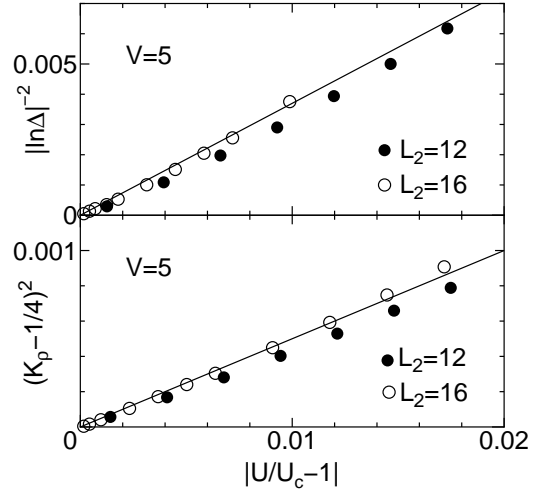}
\end{center}
  \caption[]{
$|\ln\Delta|^{-2}$  and $(K_{\rho}-1/4)^2$ as a function of $|U/U_c-1|$  near the MIT  at  $V=5$  for $L_2=12$  and $16$  together with the straight lines whose
 slope are $0.37$ for  $|\ln\Delta|^{-2}$ and $0.050$ for $(K_{\rho}-1/4)^2$.
Here, the factor of  $\Delta$ is determined by  fitting  $\Delta$ to the
 numerical result at  $U-U_c=1$. 
}
\label{fig:8}
\end{figure}
%*******************************************************************************

\section{Critical behavior near the MIT}
Now we examine the critical behavior of the renormalized $K_{\rho}$ and
 the charge gap $\Delta$ near the MIT.
In  the perturbative RG approach\cite{Nakamura,Kosterlitz}, the asymptotic behavior of  $\Delta$ is determined by the correlation length $\xi$ as $ \Delta \sim v_{\rho}/\xi$,
where  $\ln\xi \propto t^{-1/2}$ and $t=(\lambda -\lambda_c)/\lambda_c$.
Here,  $\lambda$ is a relevant parameter of a model indicating the deviation  from the  critical point $\lambda_c$. 
The critical behavior of $K_{\rho}$   is also obtained as $(K_{\rho}-1/4)^2 \propto t$  by analyzing  the  solution of the  RG equations on the fixed line at $G(\infty)=0$ with $K_{\rho}\geq 1/4$.\cite{Kosterlitz}

However, the derivation of $\xi$ in the perturbative RG approach is not clear because this approach leads the running coupling constants $G(\ell)$  and $K_{\rho}(\ell)$ into divergence in the gapfull  region with $K_{\rho}<1/4$.
Further, the perturbative RG method  fails to determine  the explicit values of the prefactors of '$t^{-1/2}$' and '$t$' in the formulation of $\ln\xi$ and  $(K_{\rho}-1/4)^2$, respectively.

On the other hand, in the combined approach of the ED and the Kehrein's RG methods, we can  explicitly determine these factors  without ambiguity.
To avoid the difficulty of the  divergence, Kehrein \cite{Kehrein1,Kehrein2} introduced a  renormalized coupling constant $\tilde{G}(\ell)$  constructed by the  product of $G(\ell)$ and the effective energy scale  $e^{-\ell(2-8K_{\rho})}$. 
In the limit $\ell \to \infty$, $G(\ell)$ diverges in proportional to
 $e^{\ell(2-8K_{\rho})}$ (see Eq. (\ref{Gell})), while $\tilde{G}(\ell)$  
 remains a  finite value and is related to the charge  gap as 
\begin{equation}
 \Delta = cv_{\rho}\tilde{G}(\infty),   
   \label{c}
\end{equation}
where $c$ is a factor of  the order of unity. Here, the explicit value $c$  is   estimated  by fitting the  RG result of $\tilde{G}(\infty)$ to the ED result of $\Delta$  and $v_{\rho}$ is calculated by the ED method.\cite{v_rho}
 The critical behavior of $\Delta$ including the explicit value of the prefactor is  determined by $\tilde{G}(\infty)$  and   
the critical behavior of $K_{\rho}$   is obtained straightforwardly from the running coupling constant $K_{\rho}(\infty)$.

\subsection{$V$-dependence for $U=\infty$}

First, we examine the critical behavior near the MIT for $U=\infty$, where 
the MIT takes place when $V$ is varied. 
In this case, we can test the reliability of our approach 
by comparing with the available exact results. 
Figure 4 shows  the critical behavior of the charge gap $\Delta$
 calculated from the combined approach of the ED and the RG methods with 
 $L_2=12$, 16 and 20  at $U=\infty$,  where  $|\ln \Delta|^{-2}$ is plotted 
 as a  function of  $(V/V_c-1)$  together with the exact result.
In the inset in Fig. 4, $\Delta$ is plotted  as a function of $V-V_c$.
Here, the factor $c$ in Eq. (\ref{c}) is determined by  fitting the RG 
result of $\Delta$ to the numerical result from the ED method at  $V-V_c=1$, 
where the system is away from the critical regime of the MIT and the ED 
method without the RG method can give an accurate result of $\Delta$. 

As shown in Fig. 4, the critical behavior from our approach agrees very well 
with the exact result which is given by 
\begin{equation}
 \frac{1}{|\ln\Delta|^2}=c_\Delta\left(\frac{V}{V_c}-1\right),  \label{c1}
\end{equation}
for $0<(V/V_c-1)\ll 1$, where $c_\Delta=\frac{8}{\pi^4}$ \cite{Yang}.  
From the results shown in Fig. 4, we estimate the coefficient as 
$c_\Delta \sim 0.087$, $0.084$ and $0.082$ for $L_2$=12, 16 and 20, respectively. Assuming the size dependence  as  $\propto 1/L_2^2$, 
  we obtain an $L_2 \to \infty$ extrapolated value  $c_\Delta \sim 0.080$ 
  which  is  close to  the exact result of   $c_\Delta=\frac{8}{\pi^4} \simeq 0.0821$. 
Thus, in contrast to the perturbative RG method, our approach gives reliable 
 and explicit estimates for the charge gap $\Delta$ with very small energy scale near  the MIT.

Fig. 5 shows  $(K_{\rho}-1/4)^2$ as a function of $(1-V/V_c)$ at 
 $U=\infty$ calculated from our approach together with  the exact result.  
In the critical regime, our result  agrees very well 
with the exact result which is given by 
\begin{equation}
   \left(K_{\rho}-\frac14\right)^2 =c_K \left(1-\frac{V}{V_c}\right),   
   \label{c2}
\end{equation}
for  $0<(1-V/V_c) \ll 1$, where $c_K=\frac{1}{8\pi^2}$. 
From the results shown in Fig. 5, we estimate the coefficient as 
 $c_K \sim 0.016$, $0.015$ and $0.014$ for $L_2=12$, $16$ 
 and $20$, respectively.
Using  $1/L_2^2$ extrapolation, we obtain the coefficient in $L_2 \to \infty$ 
 as  $c_K\sim 0.013$ which  is again close to the exact result
 $c_K=\frac{1}{8\pi^2} \simeq 0.0127$. 
This indicates that our approach  gives an accurate estimate for
 $K_{\rho}$ as well as for $\Delta$ even near the MIT beyond the usual finite 
 size  scaling for the numerical diagonalization method. 

\subsection{$V$-dependence for finite $U$}

Next, we examine the critical behavior of $\Delta$ and $K_{\rho}$  in the
 finite $U$ case. In this case, there is no available exact result. 
In  Fig. 6, we plot  $|\ln \Delta|^{-2}$ as a function of $(V/V_c-1)$ 
near the  MIT for $U=10$ and $5$. 
We find that the  critical behavior for  both $U=10$ and $5$ is the 
same as that for  $U=\infty$ given in  Eq. (\ref{c1}) except the value 
of the coefficient $c_\Delta$. 
For $U=10$,  we  estimate the coefficient as 
 $c_\Delta \sim 0.15$ and $0.15$ for $L_2=12$ and $16$, respectively, 
 which yield an $L_2 \to \infty$  extrapolated value $c_\Delta\sim 0.15$. 
For $U=5$,  the values of the coefficient are $c_\Delta\sim 0.17$ and $0.18$ 
  for $L_2$=12 and 16, respectively, resulting in  an  $L_2 \to \infty$ 
  extrapolated value  $c_\Delta \sim  0.19$.

Fig. 7 shows the critical behavior of $K_{\rho}$   for $U=10$ and $U=5$. 
We again find that the  critical behavior for  both $U=10$ and $5$ is the 
same as that for  $U=\infty$ given in  Eq. (\ref{c2}) except the value 
of the coefficient $c_K$. 
We estimate the values of $c_K$ for $U=10$ as $c_K\sim 0.022$ 
and $0.022$ for $L_2$=12 and 16,  respectively.  
For $U=5$, the values of the coefficient are $0.028$ and $0.029$ 
for $L_2$=12 and 16, respectively.
These results yield  $L_2 \to \infty$ extrapolated values $c_K \sim 0.022$ 
for $U=10$  and $c_K \sim 0.030$ for $U=5$.  

When $U$ decreases from $U=\infty$, both $c_\Delta$ and $c_K$ monotonically 
increase and become  considerably large for a suitable value of $U$ such as 
$U=10$ and $5$ as compared to the corresponding values of $c_\Delta$ and $c_K$ 
for $U=\infty$.

\subsection{$U$-dependence for finite $V$}

In the large $V$ regime ($V>U$), the MIT takes place at a critical value $U_c$ 
when $U$ is varied for a fixed value of $V$ as found in Fig. 3. 
Finally, we examine  the critical behavior in this case. 
In Fig. 8, $|\ln\Delta|^{-2}$ and $(K_{\rho}-1/4)^2$ are plotted as functions
of $|U/U_c-1|$ near the MIT for $V=5$. 
We find that the critical properties as functions of $U/U_c$ are the same as 
those as functions of $V/V_c$ given in eqs. (10) and (11) except the 
values of the coefficients: 
$|\ln \Delta|^{-2}=c'_\Delta (U/U_c-1)$ and 
$(K_{\rho}-\frac{1}{4})^2=c'_K (1-U/U_c)$ 
in the limit $U\to U_c$, respectively. 
We estimate the coefficients as 
$c'_\Delta\sim 0.30$ and $c'_K\sim 0.041$ for $L_2=12$ and 
$c'_\Delta\sim 0.33$ and $c'_K\sim 0.046$ for $L_2=16$, 
which yield $L_2\to \infty$ extrapolated values as 
$c'_\Delta\sim 0.37$ and $c'_K\sim 0.050$, respectively. 
Both of $c'_\Delta$ and $c'_K$ are considerably larger than the corresponding 
values of $c_\Delta$ and $c_K$.

\section{Summary and discussion}

In summary, we studied the critical behavior  near the MIT in  
 the one-dimensional extended Hubbard model at quarter   filling by using 
 the combined approach of the ED and the RG  methods. 
In the large $U$ regime ($U>V$), the MIT takes place at a critical value 
$V_c$ when $V$ is varied for a fixed $U$, while, in the large $V$ regime 
($V>U$), it takes place at a critical value $U_c$ when $U$ is varied for a 
fixed $V$. We examined the critical behavior near the MIT for both cases. 

In the large $U$ regime, we observed the critical behavior, 
$|\ln \Delta|^{-2}=c_\Delta (V/V_c-1)$ and 
$(K_{\rho}-\frac{1}{4})^2=c_K (1-V/V_c)$, 
where the critical value $V_c$ and the coefficients $c_\Delta$ and $c_K$ are
 functions of $U$. For $U=\infty$, the estimated values of $V_c$, 
 $c_\Delta$ and $c_K$ agree well with the exact results. 
When $U$ decreases from $U=\infty$, all of $V_c$, $c_\Delta$ and $c_K$ 
monotonically increase. Both of $c_\Delta$ and $c_K$ become  considerably 
large for a suitable value of $U$ such as $U=10$ and $5$ as compared to the 
corresponding values of $c_\Delta$ and $c_K$ for $U=\infty$. 

In the large $V$ regime, we also observed the same critical properties, 
$|\ln \Delta|^{-2}=c'_\Delta (U/U_c-1)$ and 
$(K_{\rho}-\frac{1}{4})^2=c'_K (1-U/U_c)$. 
Both of $c'_\Delta$ and $c'_K$ are considerably larger than the corresponding 
values of $c_\Delta$ and $c_K$. 
For $V\gg U$, the SC phase with $K_{\rho}>1$ appears. Near the SC phase, 
it is difficult to obtain the solution for the RG equation, 
because  the umklapp scattering is canceled by the SC fluctuation. 
To examine the critical behavior near the MIT for  $V \gg U$, 
we need an improved RG approach which includes both effects of the umklapp 
scattering and the SC fluctuation.

We also obtained the phase diagram on the $U-V$ plane and found that 
the phase boundary of the MIT and the contour lines of $\Delta$ 
and $K_\rho$ near the MIT smoothly connect between the large $U$ regime 
and the large $V$ regime. 
Although it is difficult to analyze the  critical behavior for $V \gg U$, 
there  is  no   qualitative  difference in the critical behavior. 
These results suggest  that  the nature of  the MIT is essentially 
 unchanged on the $U-V$ plane over the whole parameter regime 
 including the large $U$ and the large $V$ regimes.

In the limit $V = \infty$, electrons are  completely inhibited  to occupy
the nearest neighbor site of each other. In this case, some exact results have 
been obtained in the previous works\cite{Mila,Sano1,Penc,Nakamura}: 
the ground state energy  is always zero  for $U>U_c(=4)$ and  the charge gap 
is given by $\Delta=|U-U_c|$. 
Then, the critical behavior of $\Delta$ for $V=\infty$ is completely different 
from that for finite $V$ case obtained here. 
We think that there are two possibilities to explain this differences 
as follows: 
(1) The critical behavior near the MIT changes discontinuously at a finite $V$. 
(2) The critical region smoothly shrinks with increasing $V$ and finally 
becomes zero in the limit $V = \infty$ where the different critical 
behavior is observed. 
In any case, the critical behavior near the MIT for $V\gg U$ is interesting 
problem as the competition between the SC fluctuation and the umklapp scattering becomes important, and will be studied in the future. 

\acknowledgements

This work is partially supported by the Grant-in-Aid
for  Scientific Research from the Ministry of Education,
Culture, Sports, Science and Technology of Japan.

%--------------------------------------------------------------------


\begin{references}
\bibitem{charge-gap}
The  charge gap is determined by
$  E(L/2+1) + E(L/2-1) - 2 E(L/2),
$ where $E(L/2)$ is the total energy of the ground state for a system with
 $L/2$ electrons and  is calculated  up to 16 sites systems. 
The size dependence of $\Delta(L)$ is assumed as 
$ \Delta(L) = \Delta(\infty) + c_1/L + c_2/L^2,$
 where $c_1$ and $c_2$ are constants.
\bibitem{critical-Vc}
In the case of $U=10$,  we estimate $V_c \simeq 2.58$ and $2.57$ for 
 $L_2=12$ and 16, respectively.
For $U=5$, $V_c \simeq 3.65$ and 3.57 for  $L_2=12$ and 16, respectively.
\bibitem{v_rho}
The value of $v_{\rho}$ is determined by using the relation of $v_{\rho}=D/2K_{\rho}$.
\end{references}
\end{document}